\documentclass[prb,aps,twocolumn,draft,tightenlines,%
showpacs,floatfix,superscriptaddress]{revtex4}
\usepackage{psfig}
\usepackage{amssymb}
\usepackage{colordvi}
\begin{document}
\title{Hot-electron effect in spin dephasing in $n$-type GaAs quantum wells} 
\author{M. Q. Weng}
\author{M. W. Wu}
\thanks{Author to whom correspondence should be addressed}%
\email{mwwu@ustc.edu.cn}%
\affiliation{Hefei National Laboratory for Physical Sciences at
  Microscale, University of Science and Technology of China,
Hefei, Anhui, 230026, China}
\affiliation{Department of Physics, University of Science and
Technology of China, Hefei, Anhui, 230026, China}%
\altaffiliation{Mailing Address}
\author{L. Jiang}
\affiliation{Department of Physics, University of Science and
Technology of China, Hefei, Anhui, 230026, China}%
\date{\today}
\begin{abstract}
We perform a study of the effect of the high in-plane electric field on the
spin precession and spin dephasing due to the D'yakonov-Perel' 
mechanism in $n$-type GaAs (100) quantum wells by constructing and
numerically solving the kinetic Bloch equations. We
self-consistently include all of the scattering such as 
electron-phonon, electron-non-magnetic impurity as well as the
electron-electron Coulomb scattering in our theory and systematically
investigate how the spin precession and spin dephasing are affected
by the high electric field under various conditions. 
The hot-electron distribution functions and the 
spin correlations are calculated rigorously in our theory.
It is found that the  D'yakonov-Perel' term in the electric field provides a
non-vanishing effective magnetic field that alters the
spin  precession period.  Moreover,  spin dephasing is 
markedly affected by the electric field.
The important contribution of the electron-electron scattering to the spin
dephasing is also discussed.
\end{abstract}

\maketitle

\section{Introduction}

Spintronics is an active field which studies 
processes that manipulate  the spin degree-of-freedom of 
electrons, with the goal 
of developing new electronic devices with improved performance and new 
functionality compared to the traditional ones
which are based on precise control of the charge distribution of 
electrons.\cite{wolf,spintronics} Understanding spin dephasing is 
an important prerequisite for the realization of such devices.
As most of the semiconductor electronic devices are very small
and even a small applied voltage gives a strong electric field,
these devices usually work in the hot-electron 
condition.\cite{conwell,conwell_1972}
Therefore  understanding spin dephasing in the presence of 
a strong electric field is of particular importance to the 
spintronic application.

Recent experiments have shown 
that the electron spin lifetime is very long in $n$-type Zinc-blende
semiconductors.\cite{kikkawa1,ohno1,dzioev_2001}
In theory the spin dephasing in semiconductors 
without high electric field has been extensively studied.
Three spin dephasing mechanisms have
been proposed:\cite{meier}
the Ellit-Yaffet  mechanism\cite{yafet,elliot}
which is important in the narrow-bandgap and/or high impurity-doped
semiconductors; 
the Bir-Aronov-Pikus  mechanism\cite{bap} which is important in
the pure or $p$-type semiconductors; and  the D'yakonov-Perel'
(DP) mechanism\cite{dp} which is the main spin-dephasing mechanism in
$n$-type Zinc-blende semiconductors such as GaAs and InAs. The DP
mechanism originates from the spin-orbit interaction in crystal
without the inversion center and results in spin  splitting of the
conduction band at $k\not=  0$. This is equivalent to an effective
magnetic field acting on the spin, with its magnitude and orientation
depending on the electron wavevector. Moreover, an important 
many-body spin dephasing mechanism due to combined 
effects of the inhomogeneous broadening of the spin precession 
and the spin conserving scattering by 
irreversibly disrupting the phases between spin dipoles
has been proposed recently\cite{wu_epjb_2000} and closely
studied.\cite{wu_pss_2000,wu_jpsj_2001,wu_ssc_2002,weng_prb_2003}

The study of the effect of electric fields on electron spins
in semiconductors just begins. 
Experiments have shown that the spin polarization is not destroyed
by the strong applied electric field in transport up to a few
kV/cm.\cite{hagele_1998,sanada_2002} 
It is revealed that under right configurations the
electric field can drive the electrons to approach 
a larger spin injection length.\cite{malajovich_2001,%
spintronics,schmidt,qi_prb_2003,bournel_1999,%
saikin_2003,shen_2003,pramanik_2003a}
In Ref. \onlinecite{pramanik_2003}, the spin dephasing in quantum
wires under high electric field is studied through Monte-Carlo
simulation. 
The electric manipulation of the spin of
two-dimensional (2D) electrons through the 
Rashba\cite{ras} spin-orbital  
interactions using the in-plane AC electric field
has also been proposed.\cite{rashba_2003}
Nonetheless how the hot electron effect affects  
the spin dephasing/transport is so far not 
fully investigated. 
A complete understanding of the hot-electron effect on the
spin dephasing in $n$-type GaAs quantum wells (QW's) can 
be obtained by solving the many-body kinetic 
Bloch equations\cite{wu_prb_2000,wu_epjb_2000,wu_pss_2000}
which have been applied successfully to  studying the spin
dephasing\cite{weng_prb_2003}
and spin transport recently.\cite{weng_prb_2002}

In this paper, we use the many-body kinetic equations to study
the effect of the high electric field on the spin dephasing.
The paper is organized as following:  
In Sec.\ II we  present the model and construct the kinetic Bloch
equations. Then we  show the effect of the electric field on the
spin dephasing problem by numerically solving the kinetic equations.
In Sec.\ III(A) we first discuss how the electric field affects the
spin precession. Then we devote ourselves to 
the understanding of the effect of high electric field 
on spin dephasing under various conditions, such as at
different impurity densities; temperatures; and initial spin
polarizations. We summarize the main results in Sec.\ IV.
In Appendix A we present the effect of the
electron-electron Coulomb scattering on the spin dephasing.

\section{Model and kinetic equations}

We start our investigation of an $n$-type (100)
GaAs QW of width $a$  with its growth direction 
along the $z$-axis. An uniform electric field $\mathbf{E}$ and 
a moderate magnetic field $\mathbf{B}$ are
applied along the $x$-axis (Vogit configuration). Due to the
confinement of the QW, the 
momentum along the $z$-axis of electrons is quantized. Therefore the
electrons are characterized by a subband index $n$ and a
two-dimensional momentum $\mathbf{k}=(k_x,k_y)$, together with a
spin index $\sigma(=\pm 1/2)$. 
For simplicity, we only consider the
wells of a small width so that the separation of the subband 
energy is large enough and therefore only the lowest subband
is populated and the transition to the upper subbands is
unimportant.  It is noted that due to the so called
``runaway'' effect, the single subband model is valid only when 
the electric field is less than a few kV/cm.\cite{dmitriev_2000,dmitriev_2001}
This is because when the electric field is above the threshold value, 
electrons gain energy from the field faster than they can
dissipate it by  emitting  phonons and therefore the
transition to upper subbands becomes significant.
Consequently in the present paper we  only study the case 
with the electric field up to 1\ kV/cm which is sufficiently
large to produce the hot-electron effect.
 
For $n$-type samples, spin dephasing mainly comes from the DP
mechanism.\cite{dp,meier} With the DP term included, the Hamiltonian
of the electrons in the QW is given by:
\begin{eqnarray}
  H&=&\sum_{\mathbf{k}\sigma\sigma^{\prime}}\Big\{
      (\varepsilon_{\mathbf{k}} -
      e\mathbf{E}\cdot\mathbf{R})\delta_{\sigma
\sigma^\prime}\nonumber\\
&&\mbox{}+
      [g\mu_B\mathbf{B}+\mathbf{h}(\mathbf{k})]\cdot
      \frac{\mbox{\boldmath$\sigma$\unboldmath}
_{\sigma\sigma^{\prime}}}{2}\Big\}
    c^{\dagger}_{\mathbf{k}\sigma}c_{\mathbf{k}\sigma^{\prime}}
+H_I.
  \label{eq:Ham}
\end{eqnarray}
Here $\varepsilon_{\mathbf{k}}=\mathbf{k}^2/2m^{\ast}$ is the energy
spectrum of the electron with momentum $\mathbf{k}$ and effective mass
$m^{\ast}$.   \mbox{\boldmath$\sigma$\unboldmath} are the Pauli
matrices. $\mathbf{R}=(x,y)$ is the position. 
$\mathbf{h}(\mathbf{k})$  represents the DP term 
which serves as an effective magnetic field 
with its magnitude and direction depending
on ${\bf k}$. It is composed of 
the Dresselhaus term\cite{dress} and the Rashba term.\cite{ras}
For GaAs QW, the leading term is
the Dresselhaus term which can be written as:
\begin{eqnarray}
  \label{eq:DP}
  h_x(\mathbf{k}) &=& \gamma k_x (k_y^2-\langle k_z^2\rangle)\ ;\nonumber\\
  h_y(\mathbf{k}) &=& \gamma k_y (\langle k_z^2\rangle-k_x^2)\ ;\nonumber\\
  h_z(\mathbf{k}) &=& 0 \ .
\end{eqnarray}
Here $\langle k^2_z\rangle$ represents the average of the operator
$-({\partial\over\partial z})^2$ over the electronic state of the
lowest subband and is therefore  $(\pi/a)^2$.
$\gamma=(4/3)(m^{\ast}/m_{cv})(1/\sqrt{2m^{\ast
3}E_g})(\eta/\sqrt{1-\eta/3})$ and $\eta=\Delta/(E_g+\Delta)$, in
which $E_g$ denotes the band gap; $\Delta$ represents the spin-orbit
splitting of the valence band;
 and $m_{cv}$ is a constant close in magnitude to free
electron mass $m_0$. \cite{aronov} The Rashba term 
is proportional to the total
electric field. For narrow band-gap semiconductors such as InAs, the
Rashba term is dominant. Whereas for wide band-gap semiconductors like
GaAs, it is marginal in the regime of the applied electric
field we study. 
The interaction Hamiltonian $H_I$ is composed of the Coulomb interaction
$H_{ee}$, the electron-phonon scattering $H_{ph}$, as well as the
electron-impurity scattering $H_i$. Their expressions can be found in
textbooks.\cite{mahan,haug}  

In order to study the hot-electron  effect on spin dephasing,
we limit our system to a spacial homogeneous one 
in order to avoid the additional
complicity such as charge/spin diffusion. The
kinetic Bloch equations in such a system are constructed using the
nonequilibrium Green function method with the gradient 
expansion\cite{haug} and can be written as:
\begin{equation}
  \label{eq:Bloch}
  \dot{\rho}_{\mathbf{k},\sigma\sigma^{\prime}} -
  e\mathbf{E}\cdot\nabla_{\mathbf{k}}
  \rho_{\mathbf{k},\sigma\sigma^{\prime}}
  = \dot{\rho}_{\mathbf{k},\sigma\sigma^{\prime}}|_{\mathtt{coh}}
  +\dot{\rho}_{\mathbf{k},\sigma\sigma^{\prime}}|_{\mathtt{scatt}}\ ,
\end{equation}
where $\rho_{\mathbf{k}\sigma\sigma^{\prime}}$ represent the single
particle density matrix elements. The diagonal terms describe the electron
distribution functions $\rho_{\mathbf{k},\sigma\sigma}\equiv
f_{\mathbf{k}\sigma}$. The off-diagonal elements
$\rho_{\mathbf{k},{1\over 2}-{1\over 2}}=\rho_{\mathbf{k},-{1\over
    2}{1\over 2}}^\ast
\equiv \rho_{\mathbf{k}}$
stand for the inter-spin-band polarizations (spin 
coherence).\cite{wu_prb_2000} 
The second terms in the kinetic equations describe the 
momentum and  energy input from the  electric field $\mathbf{E}$. 
$\dot{\rho}_{\mathbf{k}\sigma\sigma^{\prime}}|_{\mathtt{coh}}$
on the right hand side of the equations
describe the coherent spin precession around
the applied magnetic field ${\bf B}$,
the effective magnetic field ${\bf h}({\bf k})$ 
from the DP term as well as the effective magnetic field from the
electron-electron interaction in the Hartree-Fock approximation.
$\dot{\rho}_{\mathbf{k}\sigma\sigma^{\prime}}|_{\mathtt{scatt}}$ 
denote the electron-impurity, the electron-phonons, as well
as the electron-electron scattering. 
The expressions of these terms are given in Appendix B.

The initial conditions at $t=0$ are taken to be
$\rho_{\mathbf{k}}(0)=0$ and electron distribution functions 
are chosen to be those of the steady 
state under the electric field but
without the magnetic field and the DP term.
Specifically $f_{\mathbf{k},\sigma}(0)$ is the steady solution of the
kinetic equations (\ref{eq:Bloch}) with 
the spin coherence $\rho_{\mathbf{k}}$, the magnetic field  and the
DP term set to be zero.
This initial distribution functions can be approached by
assuming that at time $-t_0$ there is no spin coherence
$\rho_{\mathbf{k}}(-t_0)=0$ and
the electron distributions are just the
Fermi distribution functions for each spin 
$\sigma$ at the background temperature $T$:
\begin{equation}
  f_{\mathbf{k}\sigma}(-t_0)=\Bigl\{\exp\bigl
[(\varepsilon_{\mathbf{k}}-\mu_{\sigma})/T\bigl]+1\Bigr\}^{-1}\ .
\end{equation}
and then self-consistently solving the kinetic equations
(\ref{eq:Bloch}) with the magnetic field and the DP term  turned off
(therefore no spin precession and $\rho_{\mathbf{k}}\equiv 0$).  
By taking $t_0$ to be large enough  one may get the 
steady state solution before  $t=0$. 
In Appendix A we present a typical electron distribution
function in the steady state under the electric field.
The imbalance of the the chemical potential
$\mu_{1/2}\not = \mu_{-1/2}$ gives the initial spin polarization:
\begin{equation}
  P={N_{e\;1/2}(0)-N_{e\;-1/2}(0)
    \over N_{e\;1/2}(0)+N_{e\;-1/2}(0)},
\end{equation}
where $N_{e\sigma}(t)=\sum_{\mathbf{k}}f_{k,\sigma}(t)$ is the
number of the electrons with spin $\sigma$ at time $t$. 

\section{Numerical results}

As one notices, all the unknowns to be solved
appear in the coherent and the scattering terms  
nonlinearly. Therefore the kinetic Bloch equations have to be
solved self-consistently to obtain the electron distribution and the
spin coherence.

We numerically solve the kinetic Bloch equations in such a
self-consistent fashion to 
obtain the temporal evolution of the electron distribution functions
$f_{\mathbf{k}\sigma}(t)$ and the spin coherence
$\rho_{\mathbf{k}}(t)$. Once these quantities are obtained,
we are able to deduce all the quantities such as 
electron mobility  $\mu$ and hot-electron temperature $T_e$ for
small spin polarization $P$ as well as the spin dephasing rate
for any spin polarization $P$. The mobility is given by
$\mu=\sum_{\mathbf{k}\sigma}f_{\mathbf{k}\sigma}(0)\;\mathbf{k}/N_e$;
the electron temperature is obtained by fitting the Boltzmann
tail of the  electron distribution function; whereas
the spin dephasing rate is determined by the slope of the envelope of the
incoherently summed spin coherence $\rho(t)=\sum_{\mathbf{k}}
|\rho_{\mathbf{k}}(t)|$.\cite{kuhn_1992,wu_prb_2000,wu_epjb_2000}
It is noted that the spin dephasing time obtained in this way includes
both the single-particle and the many-body spin dephasing
contributions.

\begin{table}[htbp]
  \centering
  \begin{tabular}{lllllll}
    \hline\hline
    $\kappa_\infty$ & \mbox{}\hspace{1.25cm} &
    10.8 & \mbox{}\hspace{1.25cm} &
    $\kappa_0$ & \mbox{}
    \hspace{1.25cm} & 12.9\\
    
    $\omega_0$ & & 35.4~meV & & $m^*$ & &0.067~$m_0$\\
    $\Delta$ & &0.341~eV & &$E_g$ & &1.55~eV\\
    $g$&&0.44&&&&\\
    \hline\hline
  \end{tabular}
  \caption{Material parameters used in the numerical calculations}
  \label{tab1}
\end{table}

\begin{figure}[htbp]
  \centering
  \psfig{file=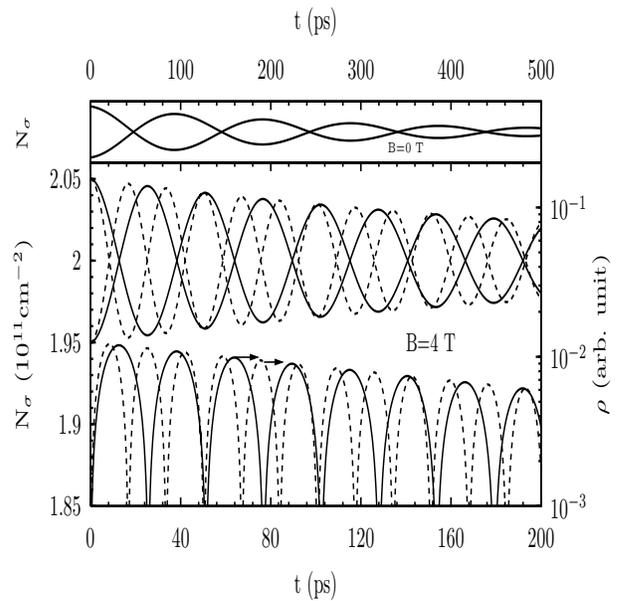,width=8.cm,height=8.cm}
  \caption{Electron densities of spin up and down and the incoherently
  summed spin coherence $\rho$ versus time $t$ for a GaAs QW with
  initial spin polarization $P=2.5$~\% under different electric fields
  $E=0.5$~kV/cm (solid) and $E=-0.5$~kV/cm (dashed).
  Top panel: $B=0$\ T; Bottom panel: $B=4$\ T.
  Note the scale of the spin coherence is on the right side of the
  figure and the scale of the top panel is different from that
of the bottom one.}
  \label{fig:tempo}
\end{figure}

We include the electron-electron, the
electron-phonon and the electron-impurity scattering
throughout our computation. For electron-phonon scattering, as we
concentrate on the relatively high temperature regime, only
electron-longitudinal optical (LO) phonon scattering is important.
The numerical scheme of the solution of the kinetic equations 
is  laid out in detail in  Appendix~\ref{appa}. 
The total electron density $N_e$, the width of
the QW $a$ and the applied magnetic field are taken to be 
$4\times 10^{11}$~cm$^{-2}$, $15$~nm and $4$~T respectively.
The material
parameters of GaAs are listed in Table~\ref{tab1}.
The numerical results are presented in Figs.~\ref{fig:tempo} 
to \ref{fig:tauP}.

\subsection{Electric field dependence of the spin precession
  frequency} 

In Fig.~\ref{fig:tempo} we plot a typical temporal evolution
of the electron densities of spin up and down for a GaAs QW with
initial spin polarization $P=2.5$\ \% under two electric fields
$E=0.5$~kV/cm and $E=-0.5$~kV/cm at
$T=120$\ K. $B=4$\ T for both cases.
The corresponding incoherently
summed spin coherence is also plotted in the figure. One can
see from the figure that, the temporal evolutions of the electron
densities and the spin coherence are similar to 
those in the absence of the applied electric
field.\cite{weng_prb_2003} The electron 
densities and the spin coherence oscillate 
as electron spins undergo the Larmor precession around the total
(effective) magnetic field. Due to the spin dephasing, the amplitude
of the oscillation decays exponentially.
An interesting effect of the high in-plane electric field on the 
spin precession is that there is marked difference in the precession 
frequency  under different electric fields (even the 
electric fields of the same magnitude but in the opposite directions). 
As shown in Fig.\ \ref{fig:tempo}, although there is almost no difference in
the corresponding spin dephasing rates,
the periods of the oscillations are
$51.2$\ ps and $33.6$\ ps for applied electric field $E=-0.5$\ kV/cm and
$0.5$\ kV/cm respectively. Both periods deviate 
from  $40.6$\ ps, which is the electric-field-free 
one of the Larmor precession under the magnetic field $B=4$\ T. 

Moreover, it is 
expected that {\em at very low} temperature ({\em i.e.}, a few Kelvin) 
where  the momentum collision rate is small, 
the DP term can result in a rapidly damped
oscillations in the spin signal when $B=0$.
At higher temperatures, due to the higher collision rates
these oscillations disappear totally
and the spin polarization decays exponentially 
with time\cite{brand_2002} and the oscillations can only be seen 
when there is an applied magnetic filed in Vogit configuration.
Nevertheless, it is of particular interest to note
in the top panel of Fig.\ \ref{fig:tempo}  that
even at temperature as high as 120\ K,  the spin signal
oscillates with period $219.9$\ ps  when there is {\em no} applied
magnetic field but an applied electric field $E=0.5$\ kV/cm. 

Both features above originate from the applied high electric field 
${\bf E}$ along the $x$-axis.
With the applied electric field, the electrons 
get a net center-of-mass drift velocity $V_d$ and the 
distribution function is no longer first-order-momentum free, {\em
  i.e.}, $\sum_{\bf k}{\bf k}f_{{\bf k}\sigma}=m^\ast V_d\not=0$.
From Eq.\ (\ref{eq:DP}) one finds that 
there is a net effective magnetic field
$B^{\ast}$ along $x$-axis from the DP term
which does not exist when $E=0$. From the period of the spin
oscillation in  Fig.\ \ref{fig:tempo}, one can deduce the effective 
magnetic field $B^\ast$. 
When electric field $E=\pm 0.5$~kV/cm, 
the net effective magnetic field $B^{\ast}=\mp 0.74$\ T.

\begin{figure}[htbp]
  \centering
  \psfig{file=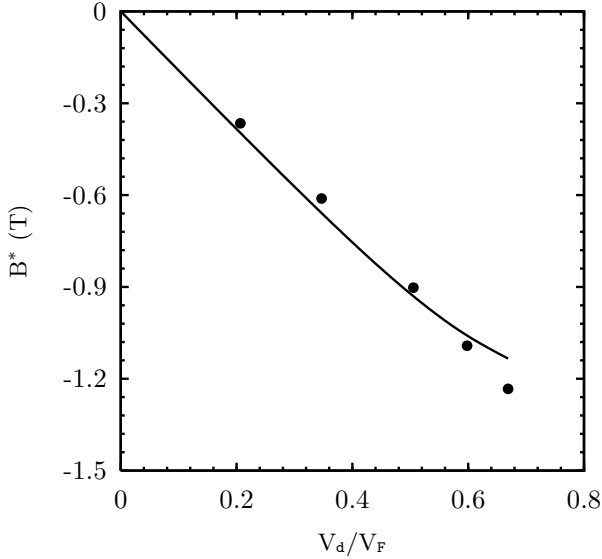,width=8.cm}
  \caption{The net effective magnetic field $B^{\ast}$ 
    from the DP term versus the drift velocity $V_d$ at $T=120$\ K 
 with impurity density $N_i=0$. The solid curve is  the corresponding
 result from Eq.\ (\ref{eq:Bdp}).
}
  \label{fig:Omg}
\end{figure}

The average of the total effective magnetic field the
electrons experience at low spin polarization 
can be given approximately by:
\begin{equation}
  \label{eq:Beff}
  \mathbf{B}_{\mathtt{tot}} =\mathbf{B}+\mathbf{B}^\ast
=\mathbf{B}+\frac{1}{g\mu_B}
\frac{\int d\mathbf{k}
    (f_{\mathbf{k}1/2}-f_{\mathbf{k}\,-{1/2}}) 
    \mathbf{h}(\mathbf{k})}
    {\int d\mathbf{k} 
    (f_{\mathbf{k}1/2}-f_{\mathbf{k}\,-1/2})}\ .
\end{equation}
By taking the electron distribution function to be the drifted Fermi
function $f_{\mathbf{k}\sigma}=\{\exp[(\mathbf{k}-m^\ast 
\mathbf{V}_d)^2/(2m^\ast)
-\mu_\sigma)/(k_BT_e)]+1\}^{-1}$, the  effective magnetic 
field for small spin polarization
can be roughly estimated as following: 
\begin{equation}
  \label{eq:Bdp}
  B^{\ast}= \gamma m^{\ast\;3}V_d 
  \bigl\{E_f/[2(1-e^{-E_f/k_BT_e})]- E_c\bigr\}/g\mu_B\,,
\end{equation}
with $E_f$ and $E_c$ standing for the Fermi energy and confinement
energy of the QW respectively. 
In Fig.\ \ref{fig:Omg}  the effective magnetic field 
$B^{\ast}$ deduced from the frequencies of  our numerical result
is plotted as a function of the drift velocity $V_d$ for the 
impurity free sample.
The result predicated by Eq.\ (\ref{eq:Bdp}) 
is  also plotted in the same figure for comparison, with the
hot-electron temperature $T_e$ obtained by fitting 
the Boltzmann tail of the calculated electron distribution functions.
It is seen
from the figure that  Eq.\ (\ref{eq:Bdp}) gives a reasonable
estimation of $B^{\ast}$. 

\subsection{Electric field dependence of the spin dephasing time of
  electrons with small spin polarization}

\begin{figure}[htbp]
  \centering
  \psfig{file=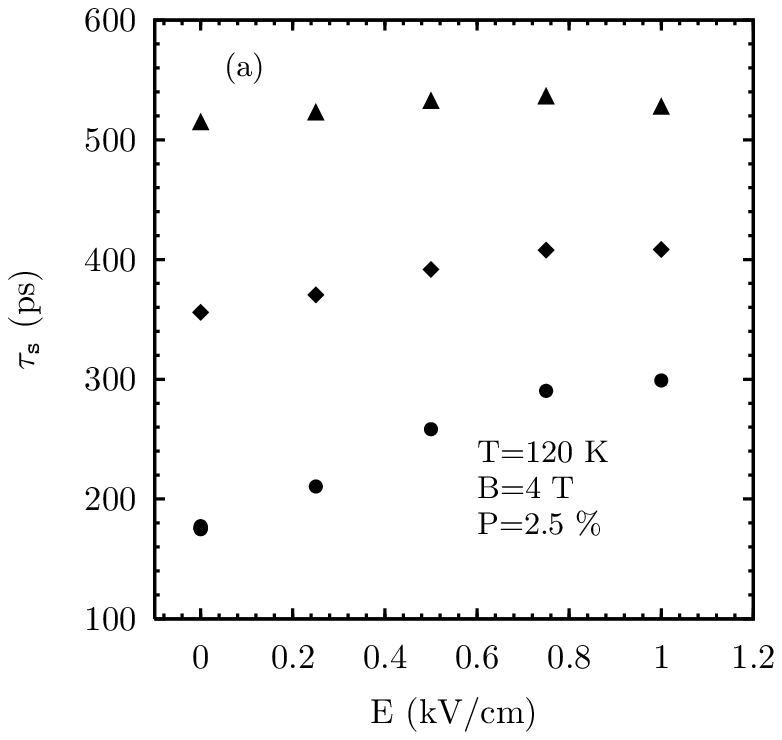,width=8.cm}
\bigskip
  \psfig{file=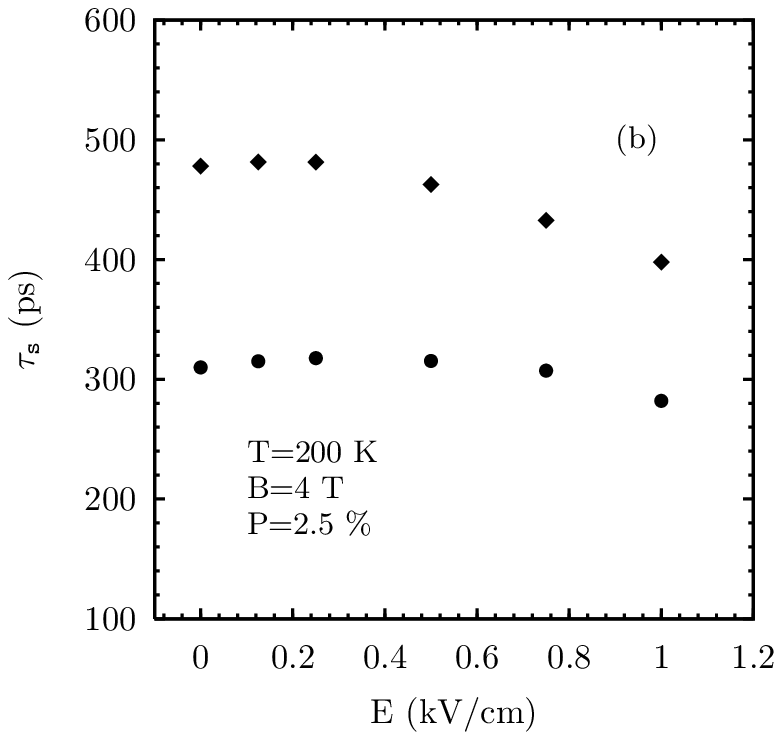,width=8.cm}
  \caption{The SDT $\tau_s$ versus the applied
    electric field $E$ at (a) $T=120$~K and (b) $T=200$~K for initial
    spin polarization $P=2.5$\ \% with
different impurity densities: $\bullet$, $N_i=0$; 
 $\blacklozenge$, $N_i=0.1\;N_e$; and
    $\blacktriangle$, $N_i=0.2\;N_e$.
  }
  \label{fig:tauE}
\end{figure}

In addition to affecting  
the spin precession frequency, the
applied high electric field also changes the spin dephasing time
(SDT), {\em i.e.} the
inverse of the spin dephasing rate, although the electric field does
not couple to the electron spin directly. 
In Fig.~\ref{fig:tauE}(a) and (b),
we plot the SDT of the electrons with initial spin polarization
$P=2.5$~\% as a function of the electric field $E$ for different
impurity densities $N_i$ at $T=120$\ K and $T=200$\ K respectively.  
It is seen from Fig.\ \ref{fig:tauE}(a) that 
for the impurity-free sample, 
the SDT first increases with the electric field 
from $\tau_s=175$\ ps at $E=0$ 
and then saturates to $\tau_s=300$~ps, $70$\ \% higher
when $E$ approaches to 1\ kV/cm.
For samples with impurities, the SDT 
also increases with the electric field but with 
decreased increase rates for higher impurity densities.
When the impurity density rises to $0.2~N_e$, $\tau_s$ first
increases for small electric field and then decreases when the
electric field is higher than $0.75$\ kV/cm. Moreover the change
of the SDT with the electric field is much smaller than 
that of the impurity-free sample. The electric field
dependence of the SDT varies as the background temperature changes. 
When the temperature is raised to a relatively high one, say
$T=200$\ K in Fig.\ \ref{fig:tauE}(b), 
the SDT increases slightly with the electric field and then
decreases when $E > 0.25$\ kV/cm 
even for the impurity-free sample. 

The electric field dependence of the SDT is understood due to
the concurrent effects of the high electric field  and
the DP term. The most obvious effect of the electric field is that the
electrons get a center-of-mass drift velocity and the center of the
distribution functions drift away from $\mathbf{k}=0$. One consequence of
the drift is that the DP term gives a net effective magnetic filed as 
discussed above. This field is moderate and hence has little effect on the SDT.
Another one is that because
more electrons are distributed at large momentum region,
the contribution from the DP term with large momentum 
is enhanced and the  SDT can be reduced. Nevertheless,
in addition to the drift, the high electric field also has 
another counter effect: As the
high electric field gives the hot-electron effect with
the electron temperature $T_e$ higher than $T$,
the scattering is strengthened. This can enhance the 
SDT.\cite{weng_prb_2003,meier}
In short, the drift of the center-of-mass in the momentum space 
trends to reduce the SDT while the hot-electron effect
helps to enhance it in the regime of our study. With these two
effects considered, the electric field dependence of the SDT 
can be understood.  

When the electric field is 
small, its effect on the DP term due to the drift is marginal.
Therefore the SDT increases with the electric field due to the
hot-electron effect when the temperature $T$ is relatively low. 
As the electric field increases, the effect of
the drift becomes important and  the SDT saturates 
consequently. It
is noted that the hot-electron effect is more pronounced
for the system with smaller impurity density
under a given electric field.\cite{lei}
As a result, the SDT increases slower with the electric field 
when the impurity density is higher.
For high impurity-doped samples, the hot-electron effect is
markedly smaller than that of the pure ones, therefore the
SDT only increases slightly in the low electric field region and then
decreases as the effect of the drift dominates. Moreover,
when the lattice temperature $T$ increases, the hot-electron effect
is also reduced. Therefore, in high temperature regime the drift
effect becomes important even for low electric fields and it becomes
possible that the SDT may drop with the increase of
the electric field even for the impurity-free  QW's. 
Moreover, the change in  $\tau_s$-$E$ curve at high temperatures
is smaller than that at low temperatures as shown in  
Fig.\ \ref{fig:tauE}.

It is noted that the electric-field dependence of $\tau_s$ we obtain
is  different from that of quantum wires where the
SDT decreases with the electric field.\cite{pramanik_2003}
This difference may come from the different contributions
of drift and hot-electron effect in quantum wells and quantum
wires. In quantum wires as the electrons  are much easier to be
accelerated by the electric field to higher momentum states.  
Therefore the drift effect is more pronounced  and it
is possible that the SDT is reduced by the electric field. While the
competing effect of the drift and the hot-electron in the QW results
in a more complicated dependence of the SDT on 
the electric field.

\begin{figure}[htbp]
  \centering
  \psfig{file=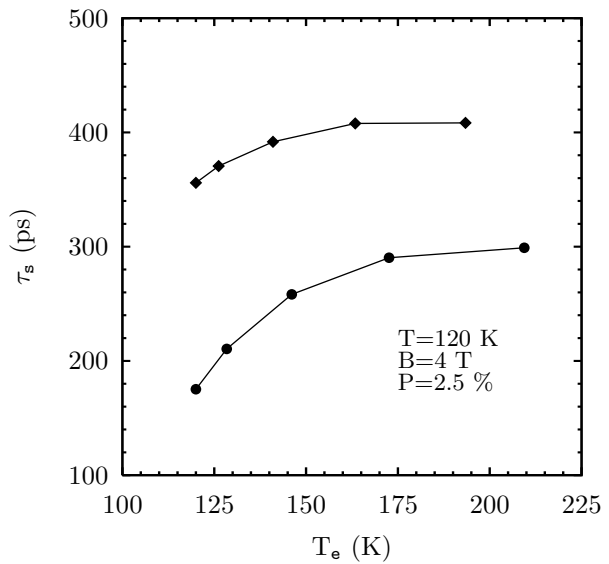,width=8.cm}
  \caption{
SDT $\tau_s$  versus the electron temperature $T_e$ at lattice
    temperature $T=120$~K 
with initial spin polarization $P=2.5$\ \%
for $N_i=0$ ($\bullet$) and $N_i=0.1\;N_e$ ($\blacklozenge$).
The curves are plotted for the aid of the eyes.
  }
  \label{fig:tauT}
\end{figure}

In order to further elucidate the effect of the high electric field to
the SDT, we replot the 
the SDT as a function of the
electron temperature $T_e$ with $T=120$~K for $N_i=0$ and
$N_i=0.1\;N_e$ in Fig.~\ref{fig:tauT}. 
It is seen that the SDT increases with the
the electron temperature $T_e$, similar to the electric-field-free case 
where the SDT increases with the
temperature.\cite{mali,weng_prb_2003}
The figure also shows that the impurities 
reduce the hot-electron effect and increase the SDT.
These results indicate that for the electric fields we study, the
electric field dependence of the SDT is affected mainly by the 
hot-electron effect in stead of the drift effect.

\subsection{Electric field dependence of the spin dephasing time of
  electrons with high spin polarization} 

\begin{figure}[htbp]
  \centering
  \psfig{file=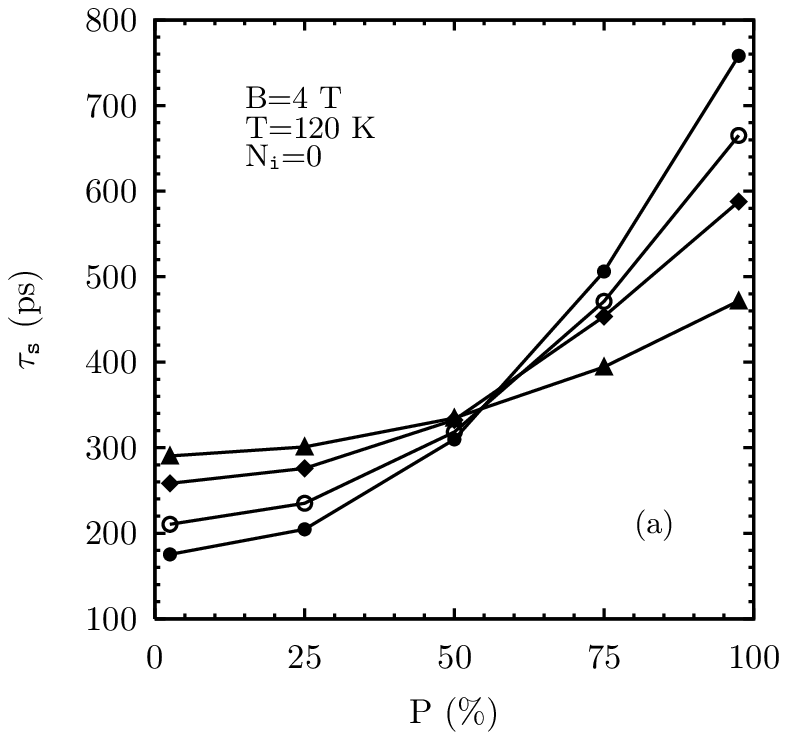,width=8.cm}
\bigskip
  \psfig{file=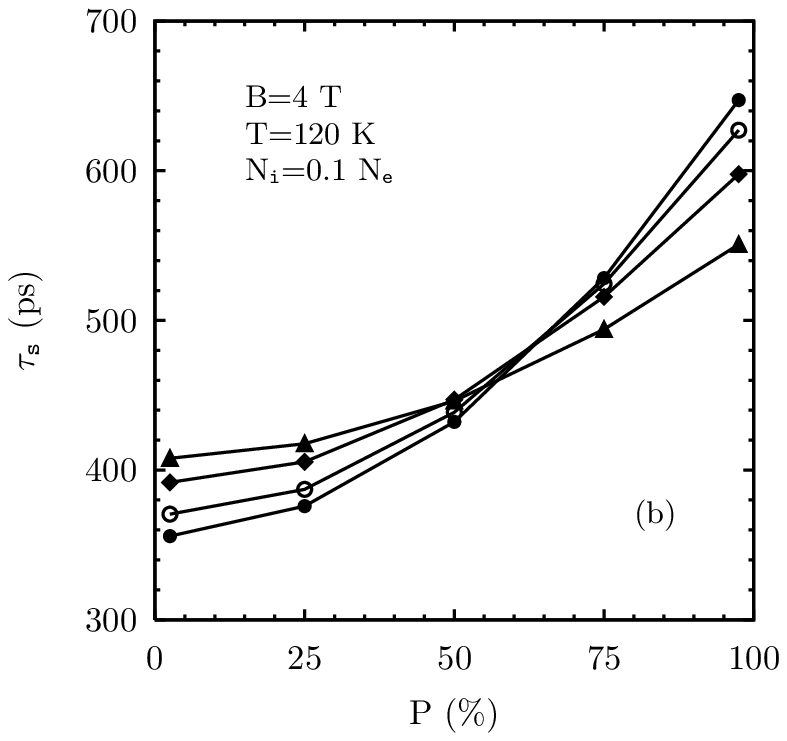,width=8.cm}
  \caption{The spin dephasing time $\tau_s$ ($\bullet$) 
    vs. the initial spin polarization for $T=120$~K and $N_i=0$ (a)
    and $N_i=0.1\;N_e$ (b) 
    under different electric field: $\bullet$, $E=0$; $\circ$,
    $E=0.25$~Kv/cm; $\blacklozenge$, $E=0.5$~Kv/cm; 
    $\blacktriangle$, $E=0.75$~Kv/cm.
  }
  \label{fig:tauP}
\end{figure}

We now turn to study the effect of the electric field on the spin
dephasing with high initial spin polarization.
Similar problem {\em in the absence} of electric fields
has been studied in our previous 
work.\cite{weng_prb_2003}

The numerical results are presented in Fig.\ \ref{fig:tauP}
where the SDT is plotted as a function of the initial spin
polarization under different electric fields. It is seen from the
figure that for all the electric fields we study, the SDT increases
with the spin polarization as the case of
$E=0$.\cite{weng_prb_2003}
However the speed drops with the increase of the electric field. 
It is interesting to see that the electric field 
dependence of the SDT is quite different for different initial spin
polarizations. In low polarization regime, the SDT increases with the
electric field while in high polarization one, it decreases
with the electric field.  For moderate spin polarized electrons, 
the SDT is insensitive to the electric field.

The rise of the SDT with the initial spin
polarization is understood due to the effective magnetic field  
from the HF term. As one component of this effective field 
is along the $z$-axis, it removes the ``detuning'' of the spin 
flip between the spin-up and -down bands and consequently
suppresses the spin  precession around  the
magnetic field and greatly reduces the
spin dephasing.\cite{weng_prb_2003} 
Therefore, all the factors, such as magnetic
field,  temperature,  impurity, electron density, and 
applied electric field which can change the HF term, 
affect the spin dephasing in the high spin polarization case
dramatically. These factors except the one of the electric field 
have been discussed in detail in our previous 
work.\cite{weng_prb_2003}
As for the factor of the applied high electric field, 
both the drift  and the hot-electron
effects affect the HF term. The drift of the center of the mass
in momentum space  provides two
competing effects on the HF term: One is to enhance the HF term
through the net effective magnetic field $B^{\ast}$
discussed above. The other is to destroy the HF term by  increasing
the DP effect. Meanwhile the hot-electron effect tends to soften the HF
term through the increase of the electron temperature and the scattering
rate.  Our results indicate that the electric field tends to
reduce the effective magnetic field from the HF term 
in high spin polarization regime and consequently 
reduce the SDT.

\section{Conclusion}

In conclusion, we have performed a systematic investigation of the
spin dephasing due to the DP mechanism in the present of high electric
fields by constructing a set of kinetic Bloch equations for
$n$-type semiconductor QW's based on the non-equilibrium Green
function method with gradient expansion. 
In our theory, we include the in-plane electric field,
the magnetic field in the Vogit configuration, the 
DP spin-orbital coupling and all the 
spin conserving scattering
such as electron-phonon, 
electron-non-magnetic impurity as
well as the electron-electron scattering. By numerically solving the
kinetic equations, we study the evolution of electron distribution
functions and the spin coherence of spin polarized electrons. The SDT
is calculated from the slope of the incoherently summed spin
coherence. In this way, we are able to study in detail how the 
spin precession and the spin dephasing are affected by the electric
field in various conditions, such as the impurity, temperature, and
spin polarization.

The in-plane electric field has two competing effects on electron spins. 
The most obvious one is that the electrons get a
center-of-mass drift velocity and the center of the distribution
functions drifts away from $\mathbf{k}=0$. One consequence of the drift
is that the DP term contributes a non-vanishing net effective magnetic
field which changes the period of the spin precession. The larger the
electric filed is, the larger the drift velocity and consequently 
the net effective magnetic field is. For the electric fields we study,
the net effective magnetic field is up to the order of 1\ T. 
This moderate magnetic field has marginal effect on the SDT
although it results in a distinct change in the spin precession
period. Another consequence  of the drift is that because more electrons are 
distributed at large momentum regime, the contribution of the DP
term with large momentum is enhanced. Therefore the drift can reduce
the SDT. In additional to the drift, the high electric field also 
introduces 
another counter effect on the spin dephasing: The scattering,
which tends to drive the electrons to the steady state, 
is enhanced as the hot-electron effect 
brought by the high electric field
with the
electron temperature $T_e$ higher than the background one $T$. That
is, the high electric field can also affect the spin dephasing through
hot-electron effect. With these two effects of the electric filed on
spin dephasing, the electric field dependence of the spin dephasing is
very rich in detail. 

In small spin polarized regime, the
hot-electron effect tends to enhance the SDT as the increase of the
scattering rate reduces the inhomogeneous broadening. Therefore in the small
electric field regime where the effect of the drift is
marginal, the SDT increases with the electric field due to the
hot-electron effect. For larger electric field, the effect of the
drift become stronger. Therefore the SDT saturates under the joint actions
of the drift and the hot-electron effect. When the impurity density or the
background temperature $T$ increases, the hot-electron effect
reduces and the effect of the drift becomes relatively
important. As a result, the increase of the SDT with the electric
field is reduced. For some regimes, the SDT decreases with the
increase of the electric field when the drift effect dominates. 

In the high spin polarized regime where the HF term plays
an important role in the spin dephasing, the hot-electron effect tends 
to reduce the SDT as both the increase of the electron temperature
$T_e$ and the increase of scattering reduce the HF term. Therefore, in
the high spin polarization regime, the SDT decreases with the increase
of the electric field. 

\begin{acknowledgments}
This work was supported by the Natural Science Foundation of China 
under Grant No. 90303012.  MWW was also supported by 
the ``100 Person Project'' of Chinese Academy of
Sciences and the Natural Science Foundation of China under Grant No. 
10247002. MQW was partially supported by China
Postdoctoral Science Foundation.
\end{acknowledgments}

\begin{figure}[htbp]
  \centering
  \psfig{file=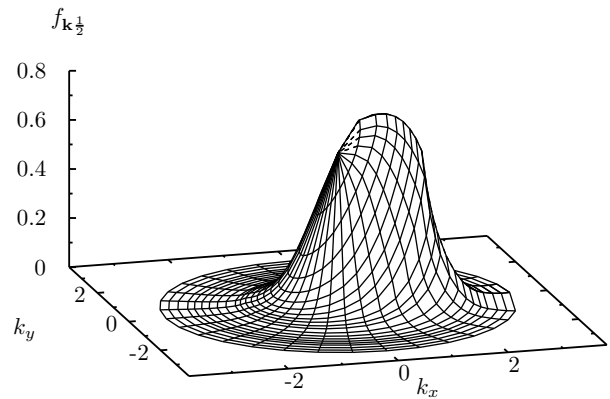,width=8.cm}
  \caption{A typical distribution function of spin-up electrons in the
steady state with electric field $E=-0.75$\ kV/cm and $P=2.5$\ \%.
  }
  \label{fig:Fk}
\end{figure}

\appendix

\section{Effect of Coulomb scattering on spin dephasing}

\begin{figure}[htbp]
  \centering
  \psfig{file=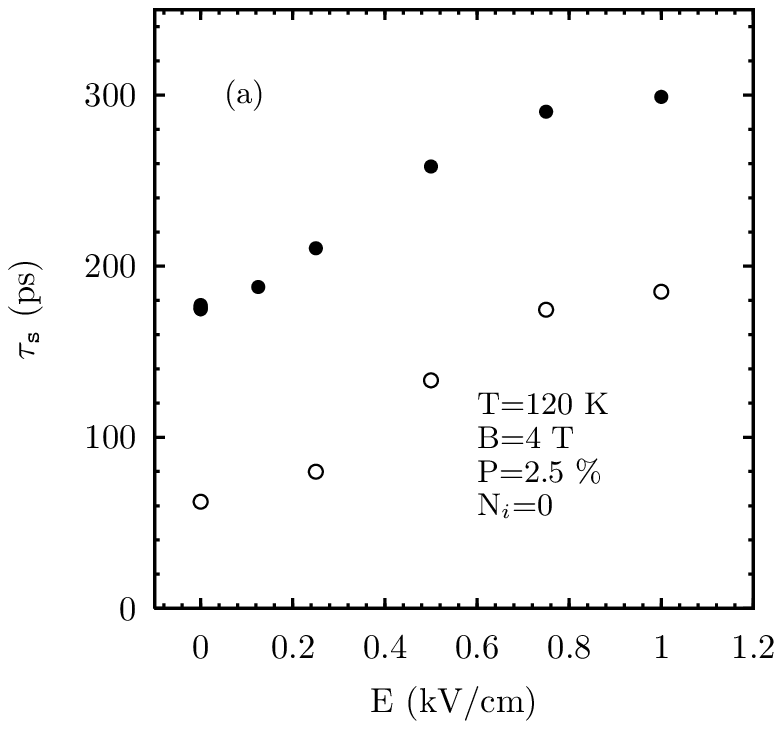,width=8.cm}
\bigskip
  \psfig{file=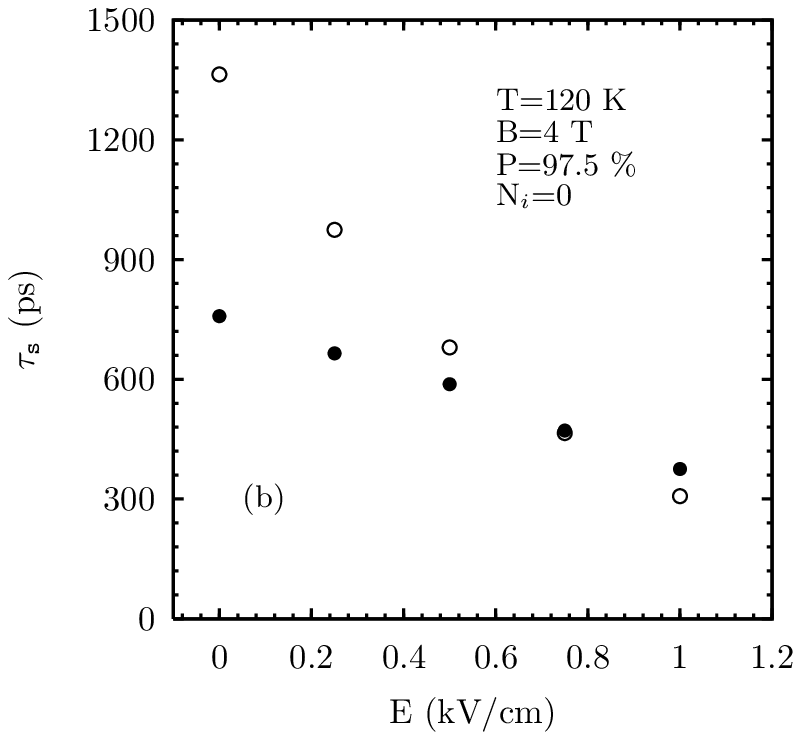,width=8.cm}
  \caption{Spin dephasing time in GaAs QW with (filled circle) and
    without (open circle) the Coulomb scattering included under
    different spin polarization: (a) P=2.5~\% and (b) P=97.5~\%.}
  \label{fig:Coul}
\end{figure}

It is stressed that the electron-electron
Coulomb scattering is of particular significance in this 
investigation. It is not only because the Coulomb scattering
is crucial in the build-up of the hot-electron temperature and
the hot-electron  distribution functions, but also because
it has strong contribution to the spin dephasing with or
without the electric field. With the Coulomb scattering, the electron
distribution functions become smoother in the momentum space and
electrons distribute more uniform around the drift center as
shown in Fig.\ \ref{fig:Fk}.

It is noted that  the treatment of the electron-electron 
scattering in the present paper takes account of the {\em full effect} of
the Coulomb scattering  which is different from our previous 
work\cite{weng_prb_2003} where the
Coulomb scattering is evaluated by replacing the 
distribution functions and the
spin coherence in the scattering 
with the corresponding isotropic averages along the angle.
In this way we are able to compute the electron distribution function
more accurate under high electric field and to have the hot-electron
effect included in our calculation. 
It is noted that in the absence of the applied electric field,
the approximation we used before greatly reduces the CPU time and 
gives good qualitative results. However, this approximation is
not good if one tries to get the results quantitatively.

We further show the importance of the Coulomb scattering in the spin
dephasing by plotting the SDT as a function of the
applied electric field with and without the Coulomb
scattering in Fig.~\ref{fig:Coul}.
 As shown in the figure, for electrons with small spin
polarization, the Coulomb scattering tends to drive the
electrons to the equilibrium state when $E=0$ or the steady state
when applied with a finite electric field, 
and hence greatly reduces the
inhomogeneous broadening originated from the DP term. 
As a result the Coulomb scattering increases the SDT 
with/without applied electric fields.
For the high spin polarized  system, as discussed before, the 
effective magnetic field along the $z$-axis from the HF term 
plays an crucial role in the spin dephasing and 
the Coulomb scattering tends to
reduce this effective magnetic field.
Therefore the SDT becomes smaller when the Coulomb
scattering is included.

\section{Numerical scheme of the kinetic Bloch equations}
\label{appa}

In this appendix we describe the scheme of the numerical solution of
the Bloch equations (\ref{eq:Bloch}). We first rewrite the
Bloch equations as following:
\begin{eqnarray}
  \dot{f}_{\mathbf{k},\sigma} &=&
  e\mathbf{E}\cdot\nabla_{\mathbf{k}}
  f_{\mathbf{k},\sigma} + 
  \dot{f}_{\mathbf{k},\sigma}|_{\mathtt{coh}}
  + \dot{f}_{\mathbf{k},\sigma}|_{\mathtt{scatt}}\ , \\
  \dot{\rho}_{\mathbf{k}} &=&
  e\mathbf{E}\cdot\nabla_{\mathbf{k}}
  \rho_{\mathbf{k}} + 
  \dot{\rho}_{\mathbf{k}}|_{\mathtt{coh}}
  + \dot{\rho}_{\mathbf{k}}|_{\mathtt{scatt}}\ .
\end{eqnarray}
The coherent terms are
\begin{widetext}
\begin{equation}
  \label{eq:f_coh}
\left. {\partial f_{{\bf k},\sigma}\over \partial t}
\right|_{\mathtt{coh}}=
-2\sigma\bigl\{[g\mu_BB+h_x({\bf k})]\mbox{Im}\rho_{{\bf k}}+h_y({\bf k})
\mbox{Re}\rho_{{\bf k}}\bigr\}
+4\sigma\mbox{Im}\sum_{{\bf q}}V_{{\bf q}}\rho^{\ast}_{{\bf k}+{\bf
    q}} \rho_{{\bf k}},
\end{equation}
\begin{eqnarray}
  \label{eq:rho_coh}
\left.{\partial \rho_{{\bf k}}\over \partial t}
\right|_{\mathtt{coh}} &=&
  {1\over 2}[ig\mu_B B + ih_x({\bf k}) + h_y({\bf k})]
  (f_{{\bf k}{1\over 2}}-f_{{\bf k}-{1\over 2}})\nonumber\\
  &&+i\sum_{{\bf q}}V_{\bf q}\bigl[(f_{{\bf k}+{\bf q}{1\over 2}}
  -f_{{\bf k}+{\bf q}-{1\over 2}})\rho_{{\bf k}}
  -\rho_{{\bf k}+{\bf q}}(f_{{\bf k}{1\over 2}}
  -f_{{\bf k}-{1\over 2}})\bigr]\ .
\end{eqnarray}
In these equations
$V_{\bf q} = \sum_{q_z}{4\pi e^2\over\kappa_0[{\bf
  q}^2+q_z^2+\kappa^2]}|I(iq_z)|^2$
with
$\kappa_0$ standing for the static dielectric constant and
$\kappa^2=6\pi N_e e^2/(a E_f)$ denoting the screening constant.
The form factor
$|I(iq_z)|^2=\pi^2\sin^2y/[y^2(y^2-\pi^2)^2]$ with $y=q_za/2$.
The scattering terms are
\begin{eqnarray}
  \label{eq:f_scatt}
\left.{\partial f_{\mathbf{k},\sigma} \over \partial t}
\right|_{\mathtt{scatt}} &=&
  \biggl\{-2\pi\sum_{\mathbf{q}q_z\lambda}g_{\mathbf{q}\lambda}^2
  \delta(\varepsilon_{\mathbf{k}}-
  \varepsilon_{\mathbf{k}-\mathbf{q}}-\Omega_{\mathbf{q}q_z\lambda})
  \bigl[N_{\mathbf{q}q_z\lambda}
  (f_{\mathbf{k}\sigma}-f_{\mathbf{k}-\mathbf{q}\sigma})
  +f_{\mathbf{k}\sigma}(1-f_{\mathbf{k}-\mathbf{q}\sigma})
  -\mbox{Re}(\rho_{\mathbf{k}}\rho^{\ast}_{\mathbf{k}-\mathbf{q}})\bigr]
  \nonumber\\
  && -2\pi N_i\sum_{\mathbf{q}}U^2_{\mathbf{q}} 
  \delta(\varepsilon_{\mathbf{k}}-\varepsilon_{\mathbf{k}-\mathbf{q}})
  \bigl[f_{\mathbf{k}\sigma}(1-f_{\mathbf{k}-\mathbf{q}\sigma})-
  \mbox{Re}(\rho_{\mathbf{k}}\rho^{\ast}_{\mathbf{k}-\mathbf{q}})\bigr]
  -2\pi\sum_{\mathbf{q}\mathbf{k}^{\prime}\sigma^{\prime}}V_{\mathbf{q}}^2 
  \delta(\varepsilon_{\mathbf{k}-\mathbf{q}}-\varepsilon_{\mathbf{k}}
  +\varepsilon_{\mathbf{k}^{\prime}}-
  \varepsilon_{\mathbf{k}^{\prime}-\mathbf{q}})
  \nonumber\\
  &&\Bigl[
  (1-f_{\mathbf{k}-\mathbf{q}\sigma})f_{\mathbf{k}\sigma}
  (1-f_{\mathbf{k}^{\prime}\sigma^{\prime}})
  f_{\mathbf{k}^{\prime}-\mathbf{q}\sigma^{\prime}}
  +{1\over 2}\rho_{\mathbf{k}}\rho^{\ast}_{\mathbf{k}-\mathbf{q}}
  (f_{\mathbf{k}^{\prime}\sigma^{\prime}}-
  f_{\mathbf{k}^{\prime}-\mathbf{q}\sigma^{\prime}})
  +{1\over 2}\rho_{\mathbf{k}^{\prime}}
  \rho^{\ast}_{\mathbf{k}^{\prime}-\mathbf{q}}
  (f_{\mathbf{k}-\mathbf{q}\sigma}-f_{\mathbf{k}\sigma})\Bigr]
  \biggr\}\nonumber\\
  &&-\{\mathbf{k}\leftrightarrow
  \mathbf{k}-\mathbf{q},\mathbf{k}^{\prime}
  \leftrightarrow\mathbf{k}^{\prime}-\mathbf{q}\},
\end{eqnarray}
\begin{eqnarray}
  \label{eq:rho_scatt}
\left.  {\partial \rho_{\mathbf{k}}\over \partial t}\right |_{\mbox{scatt}}
&=&\biggl\{
  \pi\sum_{\mathbf{q}q_z\lambda}g^2_{\mathbf{q}q_z\lambda}
  \delta(\varepsilon_{\mathbf{k}}-\varepsilon_{\mathbf{k}-\mathbf{q}}
  -\Omega_{\mathbf{q}q_z\lambda})
  \bigl[\rho_{\mathbf{k}-\mathbf{q}}
  (f_{\mathbf{k}{1\over 2}}+f_{\mathbf{k}-{1\over 2}})
  +(f_{\mathbf{k}-\mathbf{q}{1\over 2}}+f_{\mathbf{k}-\mathbf{q}-{1\over 2}}-2)
  \rho_{\mathbf{k}}
  -2N_{\mathbf{q}q_z\lambda}
  (\rho_{\mathbf{k}}-\rho_{\mathbf{k}-\mathbf{q}})\bigr] 
  \nonumber \\
  && + \pi N_i\sum_{\mathbf{q}}U_{\mathbf{q}}^2
  \delta(\varepsilon_{\mathbf{k}}-\varepsilon_{\mathbf{k}-\mathbf{q}})
  \bigl[(f_{\mathbf{k}{1\over 2}}+f_{\mathbf{k}-{1\over 2}})
  \rho_{\mathbf{k}-\mathbf{q}}
  -(2-f_{\mathbf{k}-\mathbf{q}{1\over
    2}}-f_{\mathbf{k}-\mathbf{q}-{1\over 2}})  
  \rho_{\mathbf{k}}\bigr]\nonumber\\
  &&-\sum_{\mathbf{q}\mathbf{k}^{\prime}}\pi V_{\mathbf{q}}^2
 \delta(\varepsilon_{\mathbf{k}-\mathbf{q}}-
 \varepsilon_{\mathbf{k}}+
 \varepsilon_{\mathbf{k}^{\prime}}-
 \varepsilon_{\mathbf{k}^{\prime}-\mathbf{q}})
 \biggl(
 \bigl(f_{\mathbf{k}-\mathbf{q}{1\over 2}}\rho_{\mathbf{k}}
 +\rho_{\mathbf{k}-\mathbf{q}}f_{\mathbf{k}-{1\over 2}}
 \bigr)
 (f_{\mathbf{k}^{\prime}{1\over 2}}-
 f_{\mathbf{k}^{\prime}-\mathbf{q}{1\over 2}}
 +f_{\mathbf{k}^{\prime}-{1\over 2}}-
 f_{\mathbf{k}^{\prime}-\mathbf{q}-{1\over 2}})\nonumber\\
 &&+\rho_{\mathbf{k}}\bigl[
 (1-f_{\mathbf{k}^{\prime}{1\over 2}})
 f_{\mathbf{k}^{\prime}-\mathbf{q}{1\over 2}}
 +(1-f_{\mathbf{k}^{\prime}-{1\over 2}})
 f_{\mathbf{k}^{\prime}-\mathbf{q}-{1\over 2}}
 -2\mbox{Re}(\rho^{\ast}_{\mathbf{k}^{\prime}}
 \rho_{\mathbf{k}^{\prime}-\mathbf{q}})
 \bigr]
 - \rho_{\mathbf{k}-\mathbf{q}}\bigl[
 f_{\mathbf{k}^{\prime}{1\over 2}}
 (1-f_{\mathbf{k}^{\prime}-\mathbf{q}{1\over 2}})
 \nonumber\\ &&
 +(1-f_{\mathbf{k}^{\prime}-{1\over 2}})
 f_{\mathbf{k}^{\prime}-\mathbf{q}-{1\over 2}}
 -2\mbox{Re}(\rho^{\ast}_{\mathbf{k}^{\prime}}
 \rho_{\mathbf{k}^{\prime}-\mathbf{q}})
 \bigr]\biggl)\biggl\}
 -\bigl\{\mathbf{k}\leftrightarrow \mathbf{k}-\mathbf{q},
 \mathbf{k}^{\prime}\leftrightarrow
 \mathbf{k}^{\prime}-\mathbf{q}\bigr\}\ ,
\end{eqnarray}
\end{widetext}

\begin{figure}[htbp]
  \centering
  \psfig{file=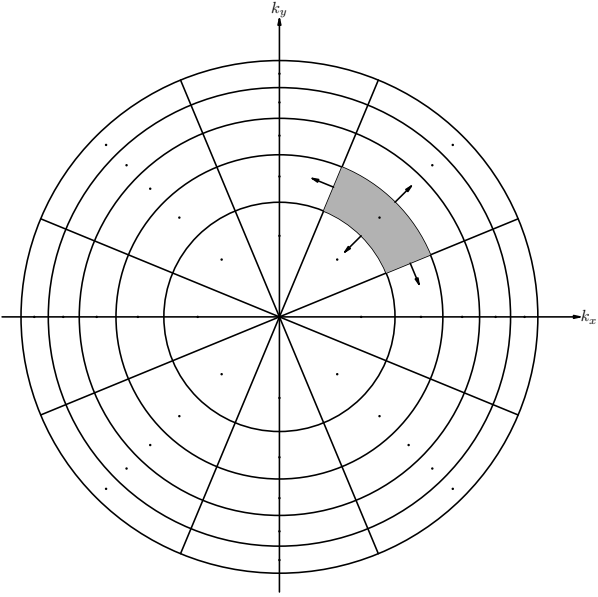,width=8.cm}
  \caption{The discrete momentum space.}
  \label{grid}
\end{figure}

\noindent in which $\{{\bf k}\leftrightarrow{\bf k}-{\bf q},
{\bf k}^{\prime}\leftrightarrow{\bf k}^{\prime}-{\bf q}\}$ stands for the
same terms as in the previous $\{\}$ but with the interchanges ${\bf
  k}\leftrightarrow {\bf k}-{\bf q}$ and ${\bf
  k}^{\prime}\leftrightarrow{\bf k}^{\prime}-{\bf q}$.
In these equations $g_{\mathbf{q},q_z,\lambda}$
are the matrix elements of electron-phonon coupling for mode $\lambda$.
For LO phonons, $g^2_{{\bf q}q_z \mbox{LO}}=
\{4\pi\alpha\Omega_{\mbox{LO}}^{3/2}/[\sqrt{2\mu}(q^2+q_z^2)]\}
|I(iq_z)|^2$ with
$\alpha=e^2\sqrt{\mu/(2\Omega_{\mbox{LO}})}
(\kappa^{-1}_{\infty}-\kappa_0^{-1})$. $\kappa_{\infty}$ is the
optical dielectric constant and $\Omega_{\mbox{LO}}$ is the LO-phonon
frequency. 
$N_{{\bf q}q_z\lambda}=1/[\exp(\Omega_{{\bf q}q_z\lambda}/k_BT)-1]$ is
the Bose distribution function of phonon with mode $\lambda$ at
temperature $T$. $U_{\bf q}^2=\sum_{q_z}\bigl\{4\pi
Z_i e^2/[\kappa_0 (q^2+q_z^2)]\bigr\}^2 |I(iq_z)|^2$ is the
electron-impurity interaction matrix element with $Z_i$ standing for the
charge number of the impurity. $Z_i$ is assumed to be $1$ throughout
our calculation.

For numerical calculation, one first 
turns the Bloch equations into discrete ones. To facilitate the
evaluation of the energy conservation, {\em i.e.} the $\delta$-function in the
scattering terms, we divide the truncated
2D momentum space into $N\times M$
control regions, each with equal energy and angle intervals as shown in
Fig.\ \ref{grid}. The $\mathbf{k}$-grid points are chosen to be
the center of each control region and therefore are written as:
\begin{equation}
  \label{eq:dis_k}
  \mathbf{k}_{n,m} = \sqrt{2m^{\ast}E_n}(\cos\theta_m, \sin\theta_m), 
\end{equation}
with $E_n=(n+1/2)\Delta E$, $\theta_m=m\Delta \theta$.
Here $n=0,1,\cdots, N-1$ and 
$m=0,1,\cdots, M-1$ with
$E_{N-1}=E_{\mathtt{cut}}$, the truncation  energy, 
and $\theta_{M-1} = (M-1)2\pi/M$. 

%

 In order to carry out the integration of the $\delta$-function
 in the scattering term, $\Delta E$ has to be chosen to satisfy 
 $\Omega_{0}=n_{\mathtt{LO}} \Delta E$ [or 
 $\Omega_{0}=(n_{\mathtt{LO}}+{1\over 2})\Delta E$]. Under this
 scheme, the coherent terms and the scattering terms of the
 electron-impurity and the electron-phonon scattering can be 
divided into discrete ones directly. Nevertheless  $f_{\mathbf{k}\sigma}$ and
$\rho_{\mathbf{k}}$ in the Coulomb scattering terms are not
  all on the grid points we choose. We approximate them to be
  the interpolation of the nearest grid points with the same energy.

The driving terms should  be treated with caution  as the equations are
stable only for some finite differential 
schemes, such as forward differencing and central differencing schemes.
In this study, we use the forward differencing scheme.
However, the usual expression of 
this scheme is based on the Taylor series expansion
and is difficult to apply to  the polar coordinate system 
which we use in this work.  
This difficulty can be circumvented by the
so called discrete conservation principle:\cite{csep} 
\begin{eqnarray}
&&\left.e\mathbf{E}\cdot\mathbf{\nabla}_{\mathbf{k}}
  f_{\mathbf{k},\sigma}\right|_{\mathbf{k}=\mathbf{k}_{n,m}} \simeq
{\int_{\Omega_{n,m}} d^2k \;
  e\mathbf{E}\cdot\mathbf{\nabla}_{\mathbf{k}}
  f_{\mathbf{k},\sigma}
  \over m^{\ast} \Delta E\Delta \theta} \nonumber \\
&& = {1\over m^{\ast} \Delta E\Delta \theta}
\int_{\partial \Omega_{n,m}} d s \; e\mathbf{E}\cdot\hat{\mathbf{n}} 
f_{\mathbf{k},\sigma} \nonumber \\
&& = {1\over m^{\ast} \Delta E\Delta \theta}
\sum_{n^{\prime}m^{\prime}}
\int_{\Omega_{n,m}\cap \Omega_{n^{\prime},m^{\prime}}} d s \; 
e\mathbf{E}\cdot\hat{\mathbf{n}} 
f_{\mathbf{k},\sigma} \nonumber \\
&& \simeq {1\over m^{\ast} \Delta E \Delta \theta} 
\sum_{n^{\prime}m^{\prime}}
e\mathbf{E}\cdot\hat{\mathbf{n}}^{n^{\prime}m^{\prime}}_{n,m}
s_{n,m}^{n^{\prime}m^{\prime}}
f_{\mathbf{k}^{\prime}_{nm},\sigma}\ . 
\end{eqnarray}
Here $\Omega_{n,m}$ and $\partial \Omega_{n,m}$ are the 
control region  which
contains the grid point $\mathbf{k}_{n,m}$ and the corresponding
boundary. In the last step of the above equation, the integration of the
boundary is replaced by the summation over the first order quadrature
on the four (or three if the control region is the neighbor of
$\mathbf{k}=0$) 
sides of the boundary $\partial\Omega_{n,m}$ with
$s^{n^{\prime}m^{\prime}}_{n,m}$ and 
$\hat{\mathbf{n}}^{n^{\prime}m^{\prime}}_{n,m}$ standing for the length 
and the outward normal to the boundary $\Omega_{nm}\cap
\Omega_{n^{\prime}m^{\prime}}$.  
In order to
satisfy the request of the numerical stability,
$\mathbf{k}^{\prime}_{n,m}$ is chosen to be $\mathbf{k}_{n,m}$ if
$-e\mathbf{E}\cdot\hat{\mathbf{n}}_{nm}^{n^{\prime}m^{\prime}} > 0$
and $\mathbf{k}_{n^{\prime},m^{\prime}}$ otherwise. 

It is noted that this choice of $\mathbf{k}$ makes our approach
identify to the 
forward differencing scheme. The time evolution is
computed by the  fourth-order Runge-Kutta 
method.\cite{numerical} 
The computation is carried out in a  parallel manner in the
``Beowulf'' cluster. For a typical calculation, it
takes about 7.5 hours to get one SDT with 16-node AMD 
Athlon XP2800$+$ CPU's when both $N$ and $M$ are chosen to be 32.

\end{document}